# Force on slowly moving variable electric or magnetic dipoles in vacuo

G. Asti and R. Coïsson
Dept. of physics and earth sciences, University of Parma
43100 Parma, Italy
e-mail giovanni.asti@fis.unipr.it and roberto.coisson@fis.unipr.it

**Abstract**
 The force on electric and magnetic dipoles moving *in vacuo* is discussed in the general case of time-variable non-uniform fields and time-variable dipole moments, to first order in *v/c* and neglecting radiation reaction. Emphasis is given to the symmetry between electric and magnetic dipoles, justifying in general Ampère's equivalence principle, and showing that the difference between gilbertian and amperian dipoles (*in vacuo*) is only a question of interpretation. The expression for the force can be expressed in a variety of different forms, and each term of each form is susceptible of specific physical interpretations. Terms not described in the literature are pointed out. A possible experiment for verifying the (*dual-Lorentz*) force of an electric field on a magnetic current (and then "*hidden momentum*") is proposed.

## 1. Introduction

In classical electrodynamics the problem of finding the general expression for the force on a stationary magnetic dipole (MD) under the action of external electric and magnetic fields has been the subject of some discussions in the literature especially in connection the question of *hidden momentum* (HM) (see [1] and referenceres therein). In particular Aharonov et al. [2], Vaidman [3] and Hnizdo [4] gave the general equation for the case of a MD of which the intensity is a function of time; and there have been discussions on the case of a moving MD in a static electromagnetic field [5,6,7]. We are here trying to give such an equation for a moving MD or electric dipole (ED), in the approximation of slow speed (i. e. neglecting terms of order $v^2/c^2$). Radiation reaction effects (due to strong and varying acceleration, high angular velocity or acceleration or fast dipole moment variation) are neglected, as is implicitly assumed in all previous literature.

Relativistic effects of first order in *v/c* appear, as is often the case of electromagnetic phenomena of moving bodies, even at extremely slow speed [1,8,9] - a subject of substantial value on the pedagogic grade, that would deserve much more attention in the practice of physics courses at educational institutions level.

We develop in parallel a description for electric and magnetic dipoles, in order to show the symmetry implied by Ampère's equivalence principle, which appears to be valid also for time-variable fields and dipoles (*in vacuo*).

We also consider a few simple examples, in particular those involving terms that have not been previously discussed in the literature. This in order to illustrate better the physical phenomena involved, and the necessity of those terms.

## 2. Forces on dipoles at rest

We summarise here the well-known results for forces on electric and magnetic dipoles at rest *in vacuo*. For MD we start using the Gilbert model (magnetic poles), in order to underline the symmetry between ED and MD, and at the end we see that the result is also valid for *amperian* dipoles.
The force on a constant electric dipole (ED) at rest in an external electromagnetic field *in vacuo* is

[10] the sum of the forces on each pole: $qE(r+h) - qE(r) \to (qh\cdot\nabla)E = (p\cdot\nabla)E$

$$F_{ed} = (p\cdot\nabla)E \tag{1e}$$

while for a magnetic dipole (MD) made of magnetic poles

$$F_{md} = (m\cdot\nabla)B \tag{1m}$$

We will call this a "gilbertian dipole", and use the same expression for the case of an ED made of electric charges.

In case the dipoles are not constant in time, another term is added [2,3,4], proportional to *dp/dt*. (or *dm/dt* for the MD)

For a physical understanding of this term we can model the electric dipole (*gilbertian* dipole), $p = qh$ with $q\to\infty$ and $h\to 0$, in two ways: two charges with constant separation with a current i flowing between the two and modifying the charges, or two constant charges with varying separation. Then in a magnetic field the internal movements cause an "internal" Lorentz force

$$f = ih\times B = \frac{dq}{dt}h\times B = \frac{dp}{dt}\times B \quad \text{or}$$

$$f = qv\times B = q\frac{dh}{dt}\times B = \frac{dp}{dt}\times B \tag{2e}$$

or, for a *gilbertian* MD, due to the *dual-Lorentz* force on the internal movement of magnetic charges $-\frac{s}{c^2}v\times E$ we get the extra force [11].

$$f = \frac{-1}{c^2}\frac{dm}{dt}\times E \tag{2m}$$

(For an amperian MD in an electric field see the discussion below about HM).

Then for time-variable dipoles

$$F_{ed} = (p\cdot\nabla)E + \frac{dp}{dt}\times B \tag{3e}$$

$$F_{md} = (m\cdot\nabla)B - \frac{1}{c^2}\frac{dm}{dt}\times E \tag{3m}$$

From equations (1) we can use the rule for the double vector product and write

$$(p\cdot\nabla)E = \nabla(p\cdot E) - p\times(\nabla\times E) \tag{4e}$$

$$(m\cdot\nabla)B = \nabla(m\cdot B) - m\times(\nabla\times B) \tag{4m}$$

and taking into account that (*in vacuo*)

$$\nabla\times B = \frac{1}{c^2}\frac{\partial E}{\partial t} \quad \text{and} \quad \nabla\times E = -\frac{\partial B}{\partial t} \tag{5}$$

we get

$$(p\cdot\nabla)E = \nabla(p\cdot E) + p\times\frac{\partial B}{\partial t} \tag{6e}$$

$$(m\cdot\nabla)B = \nabla(m\cdot B) - \frac{1}{c^2}m\times\frac{\partial E}{\partial t} \tag{6m}$$

In conclusion, adding the forces (2) due to variable *p* and *m*, we find the result found in the literature [2,3,4]:

$$F_{ed} = \nabla(p\cdot E) + \frac{d}{dt}(p\times B) \tag{7e}$$

and analogously for the magnetic dipole:

$$F_{md} = \nabla(m\cdot B) - \frac{1}{c^2}\frac{d}{dt}(m\times E) \tag{7m}$$

Here we have used total derivatives as, with dipoles at rest, partial time derivatives are also total derivatives.

This (7m) is the now well established expression (see [2-6]) for the force on an *amperian* dipole, where the second term is the time derivative of the so called "hidden momentum" (HM), a momentum due to internal interactions or movements [1].

In the ED case, eq. (7e) this momentum is part of the electromagnetic potential momentum of the field (volume integral of Poynting vector, or charges times vector potential; see Appendix 1) [12,13,14], and its time derivative can be seen as Lorentz force from the external field on internal moving charges or currents.

In eq. (7e) the second term could be as well be interpreted as due to HM, if we imagine to describe the interaction with magnetic monopoles and no electric ones, so the ED is made of a loop of circulating magnetic monopole currents, or, more realistically, consisting of a magnet rotating around one pole (we will call this an "*amperian*" ED). This term is – in practice - much bigger than the corresponding magnetic one. According to this scheme a peculiar ED could be realized by a radially magnetized cylinder in rapid rotation around its axis (something not so unrealistic, as radially magnetized bodies are commonly produced by modern permanent magnet industry).

In brief, these "momentum" terms are "external" electromagnetic momentum in the *gilbertian* model and "internal" momentum (HM) in the *amperian* model.

In conclusion, equations (7) are valid (*in vacuo*) for both gilbertian and amperian dipoles, also for time-varying dipoles and fields. The equivalence is due to the rule for the double vector product and Maxwell equations *in vacuo*.

This shows a perfect symmetry between electric and magnetic dipoles *in vacuo*, and confirms the use of Ampère's equivalence principle. In particular, to the Lorentz force on an electric current corresponds a *dual-Lorentz* force on a magnetic current.

An alternative (and simpler) derivation of eqs. (7) is sketched in Appendix 1.

**3- Force on moving dipoles**

The torque is calculated (for a constant dipole moving in a uniform constant field) in refs. [15,16].

In the dipole rest frame S', the force is given by the well-known equations (3) and (7) for a dipole at rest.

If we want to express this force as function of derivatives and fields in the lab frame S, we have to take into account that application of Lorentz transformation to dynamic field equations give rise to the following rules for the differential operators :

- the time derivative in motion in a non-uniform field is $\quad \dfrac{\partial'}{\partial t'} = \dfrac{\partial}{\partial t} + \boldsymbol{v} \cdot \nabla \quad$ (8')

- a gradient seen by an observer in motion is $\quad \nabla' = \nabla + \dfrac{\boldsymbol{v}}{c^2} \dfrac{\partial}{\partial t} \quad$ (8'')

- the dipoles see fields ($v^2/c^2 << 1$):

$$\boldsymbol{E}' = \boldsymbol{E} + \boldsymbol{v} \times \boldsymbol{B} \quad \text{and} \quad \boldsymbol{B}' = \boldsymbol{B} - \dfrac{1}{c^2} \boldsymbol{v} \times \boldsymbol{E} \quad (8''')$$

where the primed variables refer to the dipole rest system S', which is moving with velocity *v* with respect to the "lab" system S.

To first order in $v^2/c^2$, the force $\boldsymbol{F}'$ is the same as $\boldsymbol{F}$.

**3a) "*Gilbertian*" representation of the dipoles**, i.e. starting from (3)

From (8") the transformation of $p \cdot \nabla$ gives:

$$(p \cdot \nabla')E = (p \cdot \nabla)E + \frac{1}{c^2}(p \cdot v)\dot{E} \qquad (8^{IV})$$

In order to express the force $F=F'$ explicitly as a function of the variables as seen by the observer (lab frame), substituting (8) in (3e) we obtain (to first order in $v/c$):

$$F_{ed} = (p \cdot \nabla)E + \frac{1}{c^2}(p \cdot v)\dot{E} + (p \cdot \nabla)(v \times B) + \dot{p} \times B - \frac{1}{c^2}\dot{p} \times (v \times E) \qquad (9e.1)$$

(the dots indicate partial derivative with respect to time)

which, as $\dot{p} \times (v \times E) = (\dot{p} \cdot E)v - (\dot{p} \cdot v)E$ can also be written

$$F_{ed} = (p \cdot \nabla)E + v \times (p \cdot \nabla)B + \dot{p} \times B + \frac{1}{c^2}(p \cdot v)\dot{E} + \frac{1}{c^2}(\dot{p} \cdot v)E - \frac{1}{c^2}(\dot{p} \cdot E)v \qquad (9e.2)$$

The analogous case of the (*gilbertian*) MD gives then

$$F_{md} = (m \cdot \nabla)B - \frac{1}{c^2}v \times (m \cdot \nabla)E - \frac{1}{c^2}\dot{m} \times E + \frac{1}{c^2}(m \cdot v)\dot{B} + \frac{1}{c^2}(\dot{m} \cdot v)B - \frac{1}{c^2}(\dot{m} \cdot B)v \qquad (9m)$$

**3b) "A*mperian*" representation** of the dipoles. i.e. starting from (7)

The results must be the same as the starting points are equivalent, but is not obvious to show it, and this exercise can be the occasion of some physical considerations. Using (8) we obtain

$$F_{ed} = (\nabla + \frac{v}{c^2}\frac{\partial}{\partial t})(p \cdot E) + \nabla(p \cdot v \times B) + (\frac{\partial}{\partial t} + v \cdot \nabla)[p \times B - \frac{1}{c^2}p \times (v \times E)] \qquad (10e.1)$$

Developing the derivatives and neglecting terms in $v^2/c^2$, [17,18]

$$F_{ed} = \nabla(p \cdot E) + \frac{v}{c^2}(p \cdot \dot{E}) + \nabla(p \cdot v \times B) + \frac{\partial}{\partial t}(p \times B) - \frac{1}{c^2}\frac{\partial}{\partial t}[p \times (v \times E)] + (v \cdot \nabla)(p \times B) \qquad (10e.2)$$

(Note: in the expression of $\nabla'(p \cdot E')$ the gradient operator is only intended to act on the second factor, i.e. *E*. As a consequence the time derivative on *p* in the second term is absent. This condition is not evident in the expression itself, a fact that can lead sometimes to well known ambiguities [17,18]).

Also, taking into account that

$$(v \cdot \nabla)(p \times B) = p \times (v \cdot \nabla)B \text{ and } p \cdot v \times B = p \times v \cdot B \qquad (10e.2')$$

we can write

$$F_{ed} = \nabla(p \cdot E) + \frac{v}{c^2}(p \cdot \dot{E}) + \nabla(p \times v \cdot B) + \frac{\partial}{\partial t}(p \times B) - \frac{1}{c^2}\frac{\partial}{\partial t}[p \times (v \times E)] + p \times (v \cdot \nabla)B \qquad (10e.3)$$

Making use of the identity ([5], eq.13):

$$v \times (p \cdot \nabla)B = p \times (v \cdot \nabla)B + \nabla(p \times v \cdot B) \qquad (10e.3'')$$

we can also write

$$F_{ed} = \nabla(p \cdot E) + \frac{v}{c^2}(p \cdot \dot{E}) + v \times (p \cdot \nabla)B + \frac{\partial}{\partial t}(p \times B) - \frac{1}{c^2}\frac{\partial}{\partial t}[p \times (v \times E)] \qquad (10e.4)$$

We see that the third and last term in (10e.3) combine into the third term in (10e.4), whose meaning is the Lorentz force on the dipole [5]. This means that the Lorentz force comes from two contributions: one is the gradient of the scalar product of *B* with the joint magnetic dipole $p \times v$ caused by motion, and the other one comes from the (convective) time derivative of momentum . This fact happens also in the case of a single charge moving in a magnetic field [14].

We can further develop the double vector product

$$\boldsymbol{p}\times(\boldsymbol{v}\times\boldsymbol{E})=(\boldsymbol{p}\cdot\boldsymbol{E})\boldsymbol{v}-(\boldsymbol{p}\cdot\boldsymbol{v})\boldsymbol{E} \qquad (10e.4')$$

and obtain, after simplifying two terms:

$$\boldsymbol{F}_{ed}=\nabla(\boldsymbol{p}\cdot\boldsymbol{E})+\boldsymbol{v}\times(\boldsymbol{p}\cdot\nabla)\boldsymbol{B}+\frac{\partial}{\partial t}(\boldsymbol{p}\times\boldsymbol{B})+\frac{1}{c^2}[(\dot{\boldsymbol{p}}\cdot\boldsymbol{v})\boldsymbol{E}+(\boldsymbol{p}\cdot\boldsymbol{v})\dot{\boldsymbol{E}}-(\dot{\boldsymbol{p}}\cdot\boldsymbol{E})\boldsymbol{v}] \qquad (10e.6)$$

which is equivalent to (9e.2).
The magnetic equivalent is then also (9m).

**5. Simple cases and physical interpretations.**

These several expressions, as well as other ones that can be written using the various mathematical identities and/or Maxwell equations, are useful in order to analyse the physical interpretations of each term, and some of these have already been indicated in the previous section. Here we add some of the many possible remarks, as each case is susceptible of different "interpretations".
It is interesting to make simple cases where one or a few terms only survive, and this can convince us of the necessity of each term.
The most interesting examples come from the cases ($v \neq 0$) where the dipole moments and/or the fields vary with time, as these seem to imply some results not described in the literature. In order to isolate the interesting terms let us consider spatially uniform fields.

But really "simple" cases are not so simple: electromagnetic momentum does not transform as a 4-vector, as other forces and internal movements are playing a role.
Momentum density is a column (or row) of a tensor which is the sum of Maxwell stress tensor, the mechanical stress tensor and a kinetic tensor [19]. Then the momentum of an ED can be expressed in the form :

$$\boldsymbol{Q}=(\boldsymbol{p}\cdot\nabla)\boldsymbol{A}+(\boldsymbol{p}\cdot\boldsymbol{E})\boldsymbol{v}/c^2+\underline{\boldsymbol{\tau}}\cdot\boldsymbol{v}\,s\,h/c^2 \qquad (11)$$

(the second term is energy times $v/c^2$, $\underline{\boldsymbol{\tau}}$ is the mechanical stress tensor, $h$ is the dipole length and $s$ its cross-section) (we neglect the "bare" mass as it is constant in our approximation)
The second and third term are zero in the rest frame, and they appear in motion due to the effect of transformation of coordinates on the total stress tensor.
Then the mix of electromagnetic and "hidden" momentum varies when transforming from one reference system to another one. And if the dipole energy changes, the corresponding change in mass implies a change in momentum and then a force.
The following examples are an occasion for some physical considerations, but other comments are possible, and the explanations are usually hybrid, in the sense that transformation of fields, potentials, forces, velocities, times, are all intertwined.

**5.1)** ED in a uniform field $\boldsymbol{E}$ that is time-varying and perpendicular to $\boldsymbol{v}$, while $\boldsymbol{p}$ has a generic direction.

The field $\boldsymbol{E}$ is uniform in S, but due to the relativistic transformation of time (or relativity of simultaneity) it is not uniform in S', so $\quad \boldsymbol{F}=\boldsymbol{F}'=(\boldsymbol{p}\cdot\nabla')\boldsymbol{E}'=(\boldsymbol{p}\cdot\nabla)\boldsymbol{E}+\frac{1}{c^2}(\boldsymbol{p}\cdot\boldsymbol{v})\dot{\boldsymbol{E}}=\frac{1}{c^2}(\boldsymbol{p}\cdot\boldsymbol{v})\dot{\boldsymbol{E}}$

and this justifies the fifth term in (10e.6).

But this is not the only force: as the varying $\boldsymbol{E}$ implies a rotational field $\boldsymbol{B}$ and this produces a Lorentz force
$\quad \boldsymbol{F}=\boldsymbol{v}\times(\boldsymbol{p}\cdot\nabla)\boldsymbol{B}$ .

As $\quad \boldsymbol{B}=\frac{1}{2c^2}\dot{\boldsymbol{E}}\times\boldsymbol{r}$

$\boldsymbol{v}\times(\boldsymbol{p}\cdot\nabla)\boldsymbol{B}=\frac{1}{2c^2}\boldsymbol{v}\times(\boldsymbol{p}\cdot\nabla)(\dot{\boldsymbol{E}}\times\boldsymbol{r})=\frac{1}{2c^2}\boldsymbol{v}\times(\dot{\boldsymbol{E}}\times(\boldsymbol{p}\cdot\nabla)\boldsymbol{r})=\frac{1}{2c^2}\boldsymbol{v}\times(\dot{\boldsymbol{E}}\times\boldsymbol{p})=\frac{1}{2c^2}(\boldsymbol{v}\cdot\boldsymbol{p})\dot{\boldsymbol{E}}-\frac{1}{2c^2}(\boldsymbol{v}\cdot\dot{\boldsymbol{E}})\boldsymbol{p}$

In conclusion the force in our case can be written $\quad F=\dfrac{3}{2c^2}(p\cdot v)\dot{E}$ (12)

(as $\quad v\cdot\dot{E}=0\quad$) in accordance with (10e.6) (second and fifth terms).

**5.2)** Varying dipole oriented along **v** and perpendicular to a constant uniform electric field **E**.

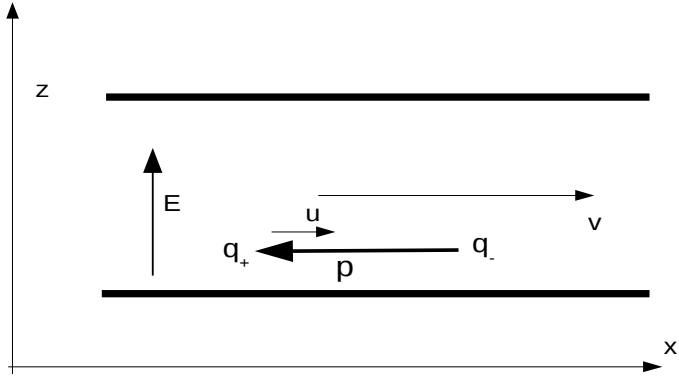

Fig.1: example illustrating the term $\dfrac{1}{c^2}(\dot{p}\cdot v)E$

An electric dipole **p=qh,** placed inside a charged plane capacitor is oriented parallel to its plates. Now we suppose that **p** is traveling at a constant speed **v** in the direction opposite to its orientation and is discharging, so that a current is flowing between its ends parallel to **v.** For simplicity let us imagine the negative charge having velocity **v** and the positive one **v+u**, with $\quad u=\dot{h}=\dot{p}/q$
Now we are asking if there is a force acting on **p.** Seemingly there no force on the dipole because it is subject to uniform electric field and its total charge is zero; moreover it is not immersed in a magnetic field, so that also the Lorentz forces are vanishing. However if we examine the system in the rest frame of the dipole, it is evident that there is active on **p** the "internal" Lorentz force
$$F'_{ed}=\dot{p}\times B'=-\dot{p}\times(v\times E)/c^2=(\dot{p}\cdot v)E/c^2=\dot{p}\,v\,E/c^2 \qquad (13)$$
And in fact this result is exactly the one that eq. (9e.2) gives us .

In S' the dipole's electromagnetic momentum is $\quad p\times B'\quad$, whose time derivative causes a force. In S this is zero.
This might be surprising, and a comment about how an e.m. force in S' can "disappear" in S is discussed in Appendix 2.

Modelling the dipole as two opposite charges connected with a rod (with variable length), the two opposite electrical forces induce a shear stress in it (which tends to induce an angular acceleration, which we consider negligible (or compensated by a suitable torque)). Moving with velocity **v** this induces a momentum density **g**, a relativistic effect due to the transformation of the stress-energy-momentum 4-tensor **τ** [19,20] ($\tau_{zx}$ being the shear stress, and in our approximation $\Upsilon=1$):
$$g_z=\dfrac{v}{c^2}\tau_{zx} \qquad (14)$$
if *qE* is the force on one pole, *h* their mutual distance, *s* the cross-section, the dipole has a
momentum $\quad Q_z=g_z s\,h=\dfrac{v}{c^2}q\,E\,h=\dfrac{-1}{c^2}p\,v\,E \qquad (15)$
(in our case the *x* component of *p* is negative)
If *p* varies, this means a force $\quad F=-\dfrac{dQ}{dt}=\dfrac{1}{c^2}\dot{p}\,v\,E \qquad (16)$
(in our case *dp/dt* is positive)

Alternatively we can say that the time of the positive charge is $\quad\delta t=t_+-t_-=-v\cdot h/c^2\quad$ in advance with respect to the negative one (*δt* is positive, as **h** is opposite to **v**).

During this time the positive charge has received an impulse, and then a momentum,
$qE\delta t = -\mathbf{p}\cdot\mathbf{v}E/c^2$ (in our case $\mathbf{p}\cdot\mathbf{v}$ is negative, then the momentum has the direction of $\mathbf{E}$, and $dp/dt$ is positive). When $p$ varies, $F = -dQ/dt = \dot{p}\,v\,E/c^2$ (16')

Lorentz transformation of the stress tensor and relativity of simultaneity are linked, as shown in another "simple case" [21].

It is interesting to remark that if we calculate the (electromagnetic) reaction force on the capacitor, this comes out equal and opposite to (16), as shown in Appendix 2.

In the dual case of a MD moving in a magnetic field, if we use the *amperian* description, the result is the homologous
$$\mathbf{F}_{md} = \dot{m}\,v\,\mathbf{B}/c^2 \tag{17}$$
but in this case in S' the momentum is "hidden", while in S the situation is similar to the electrical case, where the force is due to relativity of simultaneity and the torque.

**5.3)** Varying dipole directed along a constant and uniform field $\mathbf{E}$ and perpendicular to $\mathbf{v}$

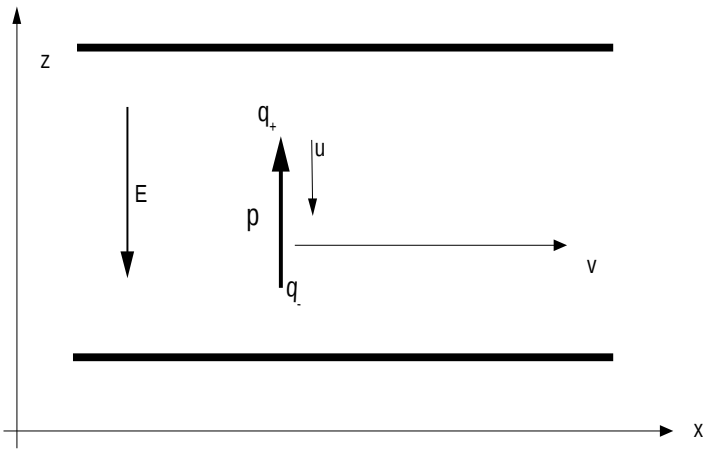

Figure 2: illustrating the term $-\dfrac{1}{c^2}(\dot{\mathbf{p}}\cdot\mathbf{E})\mathbf{v}$

We remark that $\dot{\mathbf{p}}\cdot\mathbf{E}$ is the time derivative of the energy exchanged with the field when $\mathbf{p}$ changes, and then, divided by $c^2$, of mass, and then – multiplied by $\mathbf{v}$ - the derivative of momentum, in conclusion a force directed along $\mathbf{v}$.

One may wonder why a term of the type $(\mathbf{p}\cdot\dot{\mathbf{E}})\mathbf{v}$ is not present, as it could have the same kind of explanation, but in this case there is also a rotational magnetic field.
In any case the energy balance including the internal potential energy, or current generator (in the amperian model), is delicate and might be the object of further work.

**5.4)** Case $\mathbf{p}$, $\mathbf{E}$ and $\mathbf{v}$ all parallel, with a variable $p$.
In this case our equations imply zero force ($F'=0 \rightarrow F=0$).
But there is HM, as the dipole is stretched (or compressed) by $\mathbf{E}$, and when this varies, implies a force. But when the energy $\mathbf{p}.\mathbf{E}$ varies, this implies a term $dM/dt$ ( where $M$ is the mass equivalent of energy, as in 5.3 and then a force, compensating the previous one.

## 6. A possible experiment

The force on a magnetic current in an electric field (dual-Lorentz force, as in eq.(2m)), as far as we know, has not been experimentally verified because it is very small. An experiment on a neutron beam has been proposed [22]. An experiment with a simpler apparatus to observe this force could be realised using a vibrating-wire susceptometer (VWS) [23,24,25]. The VWS, which is normally utilized for the measurement of initial magnetic susceptibility of ferromagnetic materials, consists of a thin metallic wire having the sample fixed at its centre. The wire is surrounded by a coaxial solenoid that generates an a.c. magnetic field which magnetizes the sample at half the resonance frequency of the wire. The sample is alternatively attracted by a ferromagnetic rod at its side. The measured amplitude of vibration is proportional to the apparent susceptibility of the sample. Instead in the proposed application we imagine to apply a transverse d.c. electric field, by placing an high voltage capacitor around an elongated high permeability sample, so that its magnetization change is $\delta M \approx \delta H / N$, $N$ being its demagnetizing factor in the direction of the a.c. magnetic field $\delta H = \delta H_0 \sin \omega t$.

Then the *dual-Lorentz* force we expect on the sample of volume $V$ is

$$F_m = V D \delta \dot{B} / N = \frac{V}{c^2 N} E \delta \dot{H} \ . \tag{18}$$

Let us assume the following values for the various quantities:
  $\nu = \omega/2\pi = 10^3$ Hz, $\delta H_0 = 10^3$ A/m, $E = 10^6$ V/m, $N = 10^{-2}$ $V = 10^{-9}$ m$^3$.

Then we obtain the maximum value of the force

$$F_{m0} = \frac{V}{c^2 N} E \delta H_0 2\pi\nu \ \sim 0.7 \ 10^{-11} \text{ Newtons,} \tag{19}$$

a value within the sensitivity of the VWS.

## 7. Conclusions

We have discussed the force on an electric or magnetic dipole in the general case of (rigid, i.e. not dependent on external fields) electric or magnetic dipoles (with assigned, slowly time-dependent moments) in a slowly moving ($v^2/c^2 \ll 1$, and neglecting radiation reaction) in a non-uniform time-dependent electric and magnetic field *in vacuo*. The force can be expressed in a simple and perfectly symmetric way for the electric and magnetic case, which justifies also in the dynamic case the use of Ampère's equivalence principle. The result can be expressed in a variety of forms, where each term can have a different physical interpretation.

The present discussion generalises previous results to non uniform fields, and points out extra terms in the expression of the force, illustrating the necessity of these terms. We also remark that the mix of electromagnetic and "hidden" momentum changes with the reference system. The force of the electric field on magnetic currents (or, in other words, the time derivative of HM) is – in practical situations - very small, and a possible way to experimentally observe it is suggested.

## Acknowledgements
This work has benefited from many discussions with various colleagues, in particular Vladimir Hnizdo, David Griffiths, Evgeny Bessonov, Marco Modugno and Emanuele Sorace.

## Appendix 1

The electromagnetic momentum of the electric dipole [10] is

$$\boldsymbol{Q}_{ed} = (\boldsymbol{p} \cdot \nabla) \boldsymbol{A} = \nabla (\boldsymbol{p} \cdot \boldsymbol{A}) - \boldsymbol{p} \times (\nabla \times \boldsymbol{A}) \tag{A1.1}$$

and its energy $W = \boldsymbol{p} \cdot \nabla \varphi$ , $\tag{A1.2}$

then the force is [10-12]:

$$\boldsymbol{F} = -\nabla W - d\boldsymbol{Q}/dt = -\nabla(\boldsymbol{p} \cdot \nabla \varphi) - \nabla(\boldsymbol{p} \cdot d\boldsymbol{A}/dt) + \frac{d}{dt}(\boldsymbol{p} \times \boldsymbol{B}) = \nabla(\boldsymbol{p} \cdot \boldsymbol{E}) + \frac{d}{dt}(\boldsymbol{p} \times \boldsymbol{B}) \tag{A1.3}$$

We see that the momentum in (7e) is not only in the second term, but also in a part of the first one,

and is all electromagnetic, while in (7m) the momentum is in the second term, and is all "hidden". Note also that $\nabla \phi$ and $dA/dt$ appear together, so the result depends only on $E$, independently of its origin.

## Appendix 2
Other ways to look at the "paradox" 5.2 in system S:

**(a)** Considering the forces on each charge, for ex. the negative one with velocity $v$ and the positive one $v+u$, in S' there are three forces: $qE'$ on positive charge; $-qE'$ on the negative charge; $qu' \times B'$ on the positive charge. To first order in v/c, for a particle we have $F=F'$. But in the present example we have two: one particle is moving with velocity $v$ and the other one $v+u$ ($u<<v$). While for one $F=F'$, for the other one we can write [26]:

$$F = F'(1 - \frac{1}{c^2} u' \cdot v) + \frac{1}{c^2}(u' \cdot F')v \tag{A2.1}$$

The application of this formula to the positive charge imply a reduction of the electric force exactly by the same amount that counterbalances the upward magnetic force $qu' \times B'$. This happens because it is moving away at a higher speed ($v+u > v$) with respect to the negative charge. So that the result is that the three forces add up to null, as expected. The reality is that electromagnetism alone is unable to explaining the paradox.

**(b)** The uniform constant electric field could also be obtained with two slabs parallel to x-y, each one supporting, over their surfaces, two counter-propagating uniform currents along y, producing inside the slabs two parallel and opposite magnetic fluxes, increasing with time, then producing in between a uniform linearly varying vector potential. In this case we can describe the effect in S as due to relativity of simultaneity: $\delta t = t_+ - t_- = -vh/c^2$. The interaction field momentum

$$Q = -qA(t) + q[A(t) + \frac{\partial A}{\partial t} \delta t] = -qE\frac{v}{c^2}h(t) = -p(t)E\frac{v}{c^2} \tag{A2.2}$$

and then $F = -\dot{Q} - \nabla W = \dot{p}Ev/c^2$ (remember that in our case $p$ is negative and $dp/dt$ is positive) This is what we expect, as the effect of $E$ should be the same whether it is $dA/dt$ or $grad(\varphi)$

**(c)** It is easy to see that (still in S) there is a force exerted by the dipole on the capacitor:
The discharging ED, during its motion inside the capacitor, is followed by the magnetic field that the current $i$ on the joining bar is flowing. This changing magnetic field gives rise to an electric field around the ED that has certainly an action on the charged plates of the capacitor. In order to determine such a field let us consider the following transformation equations

$$E_d = E'_d - v \times B'_d \; ; \; B_d = B'_d + v \times E'_d / c^2 \tag{A2.3}$$

where the subscript $d$ on the two fields indicates that they are generated by the moving dipole. So we see that $B_d$ generates an electrodynamic field $\quad E^* = -v \times B'_d \tag{A2.4}$

Then the elementary vertical force that this electric field causes on the lower plate is

$$dF = \sigma\, ds \cdot \left[\left(i h \times \frac{\mu_0 r}{4\pi r^3}\right) \times v\right] = i h v\, \sigma\, ds \cdot \frac{\mu_0 r}{4\pi r^3} \tag{A2.5}$$

Taking into account that the elementary area as seen from $p$ is the solid angle

$$d\Omega = ds \cdot r/r^3, \tag{A2.6}$$

we obtain the total force by integration over the positively charged plate

$$F = \int i h v \sigma\, \mathbf{ds} \cdot \frac{\mu_0 r}{4\pi r^3} = \int i h v \sigma d\Omega \frac{\mu_0}{4\pi} = -\frac{1}{2} i v h \frac{\sigma}{c^2 \varepsilon_0} = -\frac{1}{2} i h v \frac{E}{c^2} \tag{A2.7}$$

The same force is acting on the top plate thus giving the total reaction force of the ED on the capacitor $F = -\dot{p}vE/c^2$ which indeed is equal and opposite to the force acting on $p$.

# References

[1] D.J Griffiths, "Resource letter EM1-Electromagnetic Momentum", Am. J. Phys. **80**, 7-18.(2012)
[2] Y. Aharonov, P. Pearle & L. Vaidman, "Comment on "Proposed Aharonov-Casher effect: Another example of an Aharonov-Bohm effect arising from a classic lag"", Phys. Rev.A, 37, 4052-4055 (1988).
[3] L. Vaidman, "Torque and force on a magnetic dipole",  Am.J.Phys.58, 978-983 (1990)
[4] V. Hnizdo, "Hidden momentum and the force on a magnetic dipole", Magnetic and electrical separation **3**, 259-265 (1992)
[5]. V. Hnizdo, Comment on "Electromagnetic force on a moving dipole" European Journal of Physics **33**, L3-6 (2012)
[6] G.E. Vekstein "On the electromagnetic force on a moving dipole", Eur.J.Phys. **18**, 113-117 (1997); G.E. Vekstein, "Comment on 'Electromagnetic force on a moving dipole'", Eur.J.Phys. **33**, L1-L2 (2012);
[7] Kholmetskii, A. L., Missevitch, O. V., & Yarman, T. (2012). "Reply to comments on 'Electromagnetic force on a moving dipole'". European Journal of Physics, **33**(1), L7 (2011).
[8] G. Asti and  R. Coïsson "Interaction between an electric charge and a magnetic dipole of any kind (permanent, para- or dia- magnetic or superconducting)", https://arxiv.org/pdf/1506.01524v1.pdf  (2015)
[9] G. Asti "Hidden momentum in a moving capacitor",  https://arxiv.org/pdf/1508.00846v1  (2015)
[10] D. J. Griffiths "dipoles at rest" Am. J. Phys **60**, 979-987 (1992)
[11] O. Costa de Beauregard "A new law in electrodynamics", Phys. Lett. 24A, 177-178 (1967).
[12] E. J. Konopinsky  "What the electromagnetic vector potential describes," , Am. J. Phys. **46**, 499–502 (1978)
[13] M. D. Semon and J. R. Taylor "Thoughts on the magnetic vector potential", Am. J. Phys. 64, 1361 (1996);
[14] R. Coïsson "Electromagnetic interactions derived from potentials: charge and magnetic dipole" https://arxiv.org/pdf/1403.0973v1.pdf (2014)
[15] V. Namias, "Electrodynamics of moving dipoles: The case of the missing torque," Am. J. Phys. **57**, 171-177 (1989).
[16] D. J. Griffiths and V. Hnizdo, "The torque on a dipole in uniform motion",Am. J. Phys. **82**, 251–254 (2014).
[17] J.D. Jackson, "Classical Electrdynamics"(Wiley, New York, 1975) 2$^{nd}$ ed, p.185, eq 5.69
[18] T.H. Boyer, "The force on a magnetic dipole", Am. J. Phys. 56 (1988) 688
[19] C. Møller, "The theory of relativity", Oxford 1952, eq. VI-95.
[20] an effect that has been called "Poincare' stresses" in the classical model of the electron, also used to describe the Trouton-Noble effect and can be considered an example of HM in motion.
[21]  R. Coïsson and G. Guidi "Interaction momentum and simultaneity", Am. J. Phys. **69**, 492-493 (2001).
[22] J.S. Anandan "The secret life of the dipole", https://arxiv.org/pdf/hep-th/9812111v1 (1998)
[23] G. Asti and M. Solzi, "Vibrating wire magnetic susceptometer", Rev. Sci. Instrum. **67**, 3543-3552 (1996)
[24] G. Asti and M. Solzi, "A wide temperature range susceptometer", IEEE Trans Mag (1996)
[25] G. Asti, "magnetic currents and the *dual-Lorentz* force", presented at 25th Course of the International School of Materials Science and Technology, Erice 1992 (unpublished, to appear on arXiv.org)
[26] D. J. Griffiths, "Introduction to Electrodynamics", Prentice Hall, London 1999, sec. 12-2-4.